\documentclass[letterpaper, 10pt, conference]{ieeeconf}

\IEEEoverridecommandlockouts
\usepackage{times,enumerate} 
\usepackage[usenames,dvipsnames,svgnames,table]{xcolor}

\usepackage{graphicx}
\usepackage{setspace}
\usepackage{bbm}
\usepackage{mathdots,mathrsfs}
\usepackage{amssymb,latexsym,amsfonts,amsmath,cite,comment,relsize}
\usepackage{stmaryrd}
\usepackage{caption}
\usepackage[dvips]{psfrag}
\usepackage{setspace}
\usepackage{amsmath}
\usepackage{url}
\usepackage{soul}
\usepackage{blindtext}
\usepackage{graphicx}

\usepackage{epsfig} 
\usepackage{subfig}

\newtheorem{theorem}{Theorem}

\newtheorem{problem}{Problem}

\newtheorem{remark}{Remark}

\newcommand{\R}{\mathbb{R}}

\begin{document}
	
	\title{\bf Resilient Sparse Controller Design with\\ Guaranteed Disturbance Attenuation }
	
	\author{MirSaleh Bahavarnia$^\S$ and Hossein K. Mousavi$^\dagger$
		\thanks{$\S$ M. Bahavarnia is with the Department of Electrical and Computer Engineering and the Institute for Systems Research at the University of Maryland, College Park, MD 20742, USA (e-mail: mbahavar@umd.edu).}
		\thanks{$\dagger$  H.K. Mousavi is with the Department of Mechanical Engineering and Mechanics, Lehigh University, Bethlehem, PA 18015, USA (e-mail: mousavi@lehigh.edu).}
	}
	
	\IEEEoverridecommandlockouts
	
	\maketitle

	\begin{abstract}
		We design resilient sparse state-feedback controllers for a linear time-invariant (LTI) control system while attaining a pre-specified guarantee on ${\mathcal{H}}_\infty$ performance measure. We leverage a technique from non-fragile control theory to identify a region of resilient state-feedback controllers. Afterward, we explore the region to identify a sparse controller. To this end, we use two different techniques: the greedy method of sparsification, as well as the re-weighted $\ell_1$ norm minimization. Our approach highlights a tradeoff between the sparsity of the feedback gain, performance measure, and fragility of the design. To best of our knowledge, this work is the first framework providing performance guarantees for sparse feedback gain design. 
	\end{abstract}
	
	\section{Introduction}
	During the last two decades, the {robust resilient control} has been an active branch of research in control theory \cite{takahashi2000robust,famularo2000robust,park2004robust,peaucelle2005ellipsoidal}\footnote{This field of research is alternatively called robust non-fragile control.}. In this context, it is of paramount interest to design controllers that are not only robust against the exogenous processes such as disturbance or feedback noises but also resilient to perturbations in the controller space. Designing controllers with this property is undoubtedly necessary because in many real-world applications we always face uncertainties and perturbations in the controller architecture.
	
	On the other hand, {sparse control design} has amused the control theorists in recent years \cite{rotkowitz2006characterization,motee2008optimal,lin2013design,dorfler2014sparsity,dhingra2016method,arastoo2016closed,bahavarnia2017state,bahavarnia2019state,sojoudi2011structurally,lavaei2013optimal,sadabadi2014fixed,wang2014sparse,fattahi2017convexity}. In sparse control design, the main objective is to strike a balance between the performance and structure of the controller. In many cases, the goal is to push more elements of the feedback gains to be zero, while the performance of the resulting design is satisfactory. This objective may arise from a need to promote decentralization in large-scale systems wherein traditional (or classic) centralized controllers may no longer be applicable. From another point of view, sparse feedback gains reduce the computational burden in the implementation of control protocols. The research in this area has found applications, for instance in synchronization networks \cite{wu2017sparsity}, dynamic mode decomposition \cite{jovanovic2014sparsity}, and adaptive optics systems \cite{yu2018structured}.
	
	In this paper, we focus on a rather general idea to come up with a framework to find controllers that are simultaneously robust, resilient, and sparse. First, we use the tools from \cite{peaucelle2004lmi} to find a continuum of resilient controllers with a guaranteed $\mathcal{H}_{\infty}$ performance measure. Then, we uncover sparsified controllers that lie inside the specified geometric bounds. Our approach allows us to tune the level of sparsity using a scalar parameter, which highlights the tradeoff between the desired levels of sparsity and the performance and non-fragility of the feedback controller design.  To the best of our knowledge, this paper is the first research paper that provides performance guarantees on the designed sparse controller. 
	We have included several numerical experiments in this paper to support the theoretical contributions. While our approach does not guarantee a certain level of sparsity in the designed controller, the numerical examples create meaningfully sparse solutions.
	
	\noindent{\it Notations:} Throughout the paper, we adopt the following notations: all the vectors and matrices are represented by lower and upper case letters, respectively. The transpose of a matrix $X$ is denoted by $X^T$. The set of $n\times n$ symmetric matrices is denoted by $\mathbb{S}_n$. The identity matrix of appropriate size is denoted by $I$. The positive definite matrix inequality operators are shown by $\prec$ and $\succ$. The $2$-norm of a matrix is represented by $\|.\|_2$. The $\ell_0$ measure of a matrix is denoted by $\|.\|_0$, which is equal to the number of nonzero elements of the matrix. The $p$-norms of the matrices for $p\in \mathbb{Z}_{++}$ is denoted by $\|.\|_p$ where $\mathbb{Z}_{++}$ represents the set of all positive integers. 
	The element-wise Hadamard matrix product operator is shown by $\circ$.  A normal random variable with mean $a$ and covariance $\Sigma$ is denoted by $\mathcal{N}(a,\Sigma)$. The maximal eigenvalue of matrix $M \in \mathbb{S}_n$ is denoted by $\lambda_{\max}(M)$. The maximum singular value of a matrix $M$ is denoted by $\sigma_{\max}(M)$. The supremum of a set is denoted by $\sup$.
	
	\section{Problem Statement} \label{ProSt}
	
	We consider a plant with linear time-invariant (LTI) dynamics that are given by 
	\begin{align} \label{LTI} \left \{\begin{array}{l}
	\dot{x}(t) = Ax(t)+Bu(t)+B_v v(t) \vspace{2mm}  \\ 
	y(t) = C x(t)+ D_{gu} u(t)+D_{gv} v(t)  \end{array} \right .,
	\end{align}
	where $x(t) \in \mathbb{R}^n$, $u(t) \in \mathbb{R}^m$, $v(t) \in \mathbb{R}^{m_v}$, and $y(t) \in \mathbb{R}^{p}$ denote the state vector, control input, disturbance,  and output, respectively. One may choose to control the system using a static state-feedback controller 
	\begin{equation} \label{FC}
	u(t)=F x(t),
	\end{equation}
	for {state-feedback} controller\footnote{or alternatively, a feedback gain} $F \in \mathbb{R}^{m \times n}$, such that not only the closed-loop stability is achieved, but also desired performance objectives and constraints are satisfied. In this paper, we aim at finding state-feedback designs that are robust to disturbances, sparse, and resilient to the uncertainties in the implementation. In what follows, we elaborate  and formalize these objectives.
	
	\noindent{\it Robustness:}  in plants with an external disturbance, the $\mathcal{H}_\infty$ norm is a common measure to describe the quality of the disturbance attenuation in the output; that is  
	\begin{align}
	\|G(s)\|_{\mathcal{H}_{\infty}} := \sup \limits_{\omega}  ~ \sigma_{\max} \left (G(j\omega) \right ),
	\end{align}
	where $G(s) \in \R^{p \times m_v}$ is the transfer matrix from the disturbance $v$ to the output $y$\cite{zhou1996robust}. A $\gamma$-level disturbance attenuation concerns finding a controller $F$ such that 
	\begin{align}
	\|G(s)\|_{\mathcal{H}_{\infty}} \leq \gamma. 
	\end{align} 
	
	\noindent{\it Sparsity:}  at the same time, we need  controllers with the smallest number of nonzero elements, (i.e., a sparse design); the direct measure to consider this objective is the $\ell_0$ measure 
	$
	\|F\|_0.
	$
	A sparse controller design may require lower levels of computation or communication in the control system. Moreover, the privacy/security concern could become less of an issue in systems wherein the information for control action should be shared over a medium such as a cloud \cite{mital2015cloud} or sent over communication channels.

	\noindent{\it Resilience:}  in practice the implementation of any feedback controller always suffers from uncertainty to a certain extent. This might arise due to the round-off or modeling errors, actuator degradation, etc. Hence, we are generally interested in designs that are resilient (or non-fragile)  to such perturbations.  
	
	These objectives motivate us to define the following problem, wherein $\Omega_F$ is a region in the controller space, inside which, certain performance criteria hold. This set characterizes the structure of controller perturbations. 
	
	\begin{problem}[Robust Resilient Sparse State-Feedback] \label{pr:P1} For a desired performance level $\gamma>0$, find a state-feedback controller $F$, which solves   
		\begin{align*}
		\begin{array}{ll}
		\underset{F}{\textrm{minimize}}&~\|F\|_0 \vspace{2mm} \\
		{\textrm{subject to:}}&~ \|G(s)\|_{\mathcal{H}_{\infty}} \le \gamma, \textrm{ ~~for any } \hat F \in \Omega_F. 
		\end{array}
		\end{align*}
	\end{problem}
	
	\vspace{4mm}
	\noindent{\it Paper Objectives:}  even in the absence of resilience requirement, (i.e., the set is a singleton $\Omega_F \equiv \{F\}$), it can be shown that Problem \ref{pr:P1} is NP-hard \cite{blondel1997np}, so {we will not seek the exact solution to this problem}. Therefore, as a relaxation, we  propose a two-level algorithm to find a feedback gain $F$ such that \vspace{1mm}
	
	\noindent \textit{(i)} the controller $F$ is (hopefully) sparse. \\
	\noindent \textit{(ii)} the $\gamma$-level performance of the system is guaranteed. \\
	\noindent \textit{(iii)} we may evaluate  an explicitly defined region $\Omega_F$  in the controller space such that $F \in {\Omega_F}$ and that for any feedback gain $\hat F \in \Omega_F$, the stability and $\gamma$-level performance of the controller are still guaranteed. 
	
	In this paper, we develop a general two-stage method in which, first the region $\Omega_F$ is identified using linear matrix inequalities. Then,  a sparse controller is found inside this region in the second stage. To realize the second step, we develop two different ways that the sparsification may be conducted: first, we leverage re-weighted $\ell_1$ minimization techniques, which also result in a set of linear matrix inequalities.  Alternatively, we show that we can efficiently sparsify the controller by a greedy method. In both cases, we illustrate and control the tradeoff between resilience and sparsity of the controllers using an appropriate design parameter. 
	\section{Generic Two-Level Algorithm}
	
	First, we recall the machinery developed by Peacelle et. al. \cite{peaucelle2005ellipsoidal}, which characterizes a quadratic region of controllers inside which every controller provides us with  $\gamma$-level performance attenuation. Then, we define a generic two-level algorithm to show that how we can use the output of that method to find sparse state-feedback controllers that enjoy the robustness and resilience of the first method. 
	
	\subsection{Quadratically Resilient Robust Control}
	
	In the following theorem, we recall an important result from \cite{peaucelle2004lmi}, wherein the authors introduce a method to find a region in the controller space such that any controller lying in the region induces the $\gamma$-level disturbance attenuation.  
	
	\begin{theorem}[see \cite{peaucelle2004lmi}] \label{Theometh}
		the linear time-invariant control system \eqref{LTI} is stabilizable by means of a state-feedback controller $F$ such that the transfer matrix from the disturbance to the output satisfies $\|G(s) \|_{\mathcal{H}_{\infty}} \le \gamma$ if and only if there exists a Lyapunov matrix $P$ and three matrices $\hat{X} \in \mathbb{S}_m$, $\hat{Y} \in \mathbb{R}^{m \times n}$, and $\hat{Z} \in \mathbb{S}_{n}$, which  satisfy the linear matrix inequalities given by 
		\begin{align} \label{LMS}
		&~ \hat{X}  \preceq 0,  ~~~~P  \succ 0, ~~~~~~ \hat{Z}  \succ 0,  \\
		& \notag \begin{bmatrix}  
		Q_{11} & Q_{12}\\
		Q_{21} & Q_{22}
		\end{bmatrix}+ \begin{bmatrix}
		B_v\\
		D_{gv}
		\end{bmatrix} \begin{bmatrix}
		B_v\\
		D_{gv}
		\end{bmatrix}^T  \prec 0,
		\end{align}
		where the blocks are defined to be 
		\begin{align*}
		Q_{11}&:= AP-B\hat{Y}+\left (AP-B\hat{Y} \right )^T-B\hat{X}B^T+\hat{Z},\\
		Q_{12}&:= \left (C P-D_{gu}\hat{X}B^T-D_{gu}\hat{Y}\right )^T=Q_{21}^T,\\
		Q_{22}&:= -\gamma^2 I-D_{gu}\hat{X}D_{gu}^T.
		\end{align*}
		Moreover, for any feasible solution, if we define
		\begin{align*}
		& X := \hat{X} + \hat{Y} \hat{Z}^{-1} \hat{Y}^T,~ Y := \hat{Y} \hat{Z}^{-1} P, ~ Z := P \hat{Z}^{-1} P, \\
		& F_o := -YZ^{-1}=-\hat Y P^{-1},~ R :=YZY^T-X, 
		\end{align*}
		then, for any controller $F$ satisfying the quadratic inequality 
		\begin{align} \label{Eradius}
		(F-F_o)\,Z\,(F-F_o)^T\, \preceq\, {R},
		\end{align}
		the closed-loop systems is asymptotically stable and the performance measure of the system satisfies $\|G(s)\|_{\mathcal{H}_{\infty}} \leq \gamma$. 
	\end{theorem}
	
	We observe that the feedback gain $F_o$ is the center of the ellipsoidal region defined by \eqref{Eradius}. In fact, the quadratic constraint \eqref{Eradius} implies that all controllers that are close enough to control gain $F_o$ will inherit the stability and performance guarantee, (i.e., bound on $\mathcal{H}_\infty$ performance measure) from central controller $F_o$. 
	\subsection{Generic Search for Sparse Feedback Controllers} 
	
	Here, we go over the idea for finding sparse controllers. Suppose that we have found a quadratic resilient region; i.e., the interior of the matrix ellipsoid centered at $F_o$ defined by quadratic inequality (\ref{Eradius}). Now, we wish to find 
	a sparse controller $F$ inside this region. As long as we do not leave this region, we would benefit from the performance guarantee of  $\left \|G(s) \right \|_{\mathcal{H}_{\infty}} \le \gamma$. Moreover, if we maintain a "safe distance" from the boundary of this matrix ellipsoid, we can  preserve the resilience of the design to certain extent. To do so, instead of searching for  sparse controllers in the whole region defined by \eqref{Eradius}, we introduce a design parameter $\theta \in [0,1]$ and use it to control the breadth of the search space around the initial controller $F_o$. Mathematically, compared to the region defined by \eqref{Eradius}, we consider a possibly shrunk region of controllers 
	\begin{align} \label{eq:rad_theta}
	(F-F_o)Z(F-F_o)^T \preceq \theta {R},
	\end{align}
	for design parameter $\theta \in [0,1]$. We are interested in rewriting this inequality in a form that is linear in its argument $F$. This is particularly useful for arriving at semi-definite programs \cite{grant2008cvx}. To achieve this, one way is to apply the Schur complement \cite{zhang2006schur} on matrix inequality (\ref{eq:rad_theta}) to obtain an equivalent linear matrix inequality 
	\begin{align}
	\begin{bmatrix}
	\theta {R} & F-F_o\\
	(F-F_o)^T & Z^{-1}
	\end{bmatrix} \succeq 0. \label{Schured}
	\end{align}
	Given this search space, we define the following generic problem for finding sparse feedback gains. 
	
	\begin{problem}\label{pr:gen} Given  matrix ellipsoid (\ref{Eradius}), find a sparse controller from the optimization problem 
		\begin{align*}
		\begin{array}{ll}
		\underset{F}{\textrm{minimize}}&\|F\|_0 \vspace{2mm} \\
		{\textrm{subject to:}}& \begin{bmatrix}
		\theta {R} & F-F_o\\
		(F-F_o)^T & Z^{-1}
		\end{bmatrix}\succeq 0.  \notag
		\end{array}
		\end{align*}
	\end{problem}
	\vspace{2mm}
	
	Depending on the value of $\theta$, Problem \ref{pr:gen} will imply either of the following designs with  guaranteed $\mathcal{H}_{\infty}$ performance:
	\begin{enumerate}
		\item $\theta=0$:  quadratically resilient design, i.e., $F=F_o$.
		\item $ \theta \in (0,1)$: resilient sparse design. 
		\item $\theta=1$: sparse design. 
	\end{enumerate}
	
	We merge these two steps to form the following \emph{generic} two-stage algorithm:
	
	\begin{samepage}
		\noindent{\rule{8.65cm}{0.8pt}} \\
		{\bf Algorithm 1: \\ Generic Two-Stage Algorithm of Resilient Sparse State-Feedback Controller with  Guaranteed $\mathcal{H}_{\infty}$ Performance}
		\begin{enumerate}
			\item Find $Z,~R$, $F_o$  from  linear matrix inequalities \eqref{LMS}. 
			\item Solve Problem \ref{pr:gen} for a sparse controller $F$. 
		\end{enumerate}
		\noindent{\rule{8.65cm}{0.8pt}} 
	\end{samepage}
	
	We can call this method the generic algorithm for finding a sparse controller because at this point we have not stated how to perform the second step in Problem \ref{pr:gen}. In the next section, we preserve the constraint of Problem \ref{pr:gen} and consider tractable relaxations for the objective function, which hopefully give us controllers that are sparse. 
	
	\section{Realization of Sparsity}
	
	We consider two methods to fill in the blanks of Algorithm $1$, i.e., methods that let us find sparse feedback gains. 
	
	\subsection{Re-Weighted $\ell_1$ Regularization}
	Instead of dealing directly with Problem \ref{pr:gen}, we can consider a weighted $\ell_1$ regularization and define a problem:
	
	\begin{problem}\label{pr:reweighted} Given the matrix ellipsoid (\ref{Eradius}), find a sparse controller from the optimization problem, \begin{align*}
		\begin{array}{ll}
		\underset{F}{\textrm{minimize}}&\|W \circ F\|_1 \vspace{2mm}  \\ 
		{\textrm{subject to:}}& \begin{bmatrix}
		\theta {R} & F-F_o\\
		(F-F_o)^T & Z^{-1}
		\end{bmatrix} \succeq 0, \notag
		\end{array}
		\end{align*}
		where $W \in \mathbb{R}^{m \times n}$ represents the element-wise positive weight matrix.
	\end{problem}
	
	In fact,  the $\ell_1$ term in Problem \ref{pr:reweighted} is expected to act as a \emph{proxy} to the sparsity \cite{lin2013design}. 
	Once we relax Problem \ref{pr:gen} and replace it with Problem \ref{pr:reweighted}, we get Algorithm $2$.   
	
	\begin{samepage}
		\noindent{\rule{8.65cm}{0.8pt}} \\
		{\bf  Algorithm 2: \\
			Resilient Sparse State-Feedback Controller Design with $\mathcal{H}_{\infty}$ Performance using $\ell_1$ Regularization}
		\begin{enumerate}
			\item Find $Z,~R$, $F_o$  from the linear matrix inequalities \eqref{LMS}.  
			\item Solve Problem \ref{pr:reweighted} for a sparse controller $F$. 
		\end{enumerate}
		\rule{8.65cm}{0.8pt}
	\end{samepage}
	
	\noindent{\it Regularization with Re-weighting:} The choice of the weight matrix $W$ plays a significant role in the properties of the proposed method. When an appropriate weight matrix is not available, the re-weighted $\ell_1$ norm technique can also be utilized. In this  method, the weight assigned to each element is updated iteratively. This {element-wise} update is inversely proportional to the {absolute} value of the corresponding element in $F^{(k)}$ that is recovered from the past iteration as 
	\begin{align}
	&W_{ij}^{(k+1)}\,{\leftarrow }\,\frac{1}{|F_{ij}^{(k)}|+\zeta}, \label{eq:w_update}
	\end{align}
	where the constant $\zeta > 0$  is added to the denominator of update law (\ref{eq:w_update}) for numerical stability \cite{Candes:2008}. The value of $\zeta$ is usually chosen to be relatively small. 
	The stopping criterion is implemented by examining the ratio
	\begin{equation} \label{eq:stop_crit}
	\epsilon^{(k+1)}\,{\leftarrow}\,\frac{\|F^{(k+1)}-F^{(k)}\|_2}{\|F^{(k+1)}\|_2},
	\end{equation}
	for the $k$'th iteration. For a desired level of precision $\epsilon_d$, once $\epsilon^{(k+1)}\le \epsilon_d$ holds, we terminate the iterations. In the last step of the algorithm, we truncate the elements with negligible magnitude, e.g., those smaller than a certain threshold  (for instance, $5 \times 10^{-5}$) in   resulting controller $F$. 
	
	\begin{remark}Our simulation results are obtained by incorporating  update law \eqref{eq:w_update} for the first few iterations. Also, $W \in \R^{m\times n}$ is initially set to the matrix of all ones. 
	\end{remark}
	
	\subsection{Greedy Sparsification} 
	
	Instead of application of the regularization in the second step, we may alternatively consider the following greedy algorithm. We start with $F^{(0)}=F_o$, and iteratively for $k=1,2,\dots$, the next controller $F^{(k)}$ is derived by setting exactly one nonzero element of $F^{(k-1)}$ equal to  zero. This nonzero element is chosen such that: the \emph{distance} of   updated constraint \eqref{Schured} from the boundary of positive-definiteness  has a minimal decrease. We define our proxy to distance as follows: Let us  define the matrix 
	\begin{align}
	E^{(k)}:=
	\begin{bmatrix}
	\theta {R} & F^{(k)}-F_o\\
	(F^{(k)}-F_o)^T & Z^{-1}
	\end{bmatrix},
	\end{align}
	Also, we define  $\Delta({i,j}) \in {R^{m \times n}}$ to be a matrix of all zeros expect with a $1$ at row $i$ and column $j$. 
	Now, if we choose to push the nonzero element $F_{ij}^{(k-1)}$ to $0$, then one inspects that matrix $E^{(k)}$ is updated according to  
	\begin{align}\label{eq:firstupdate}
	E^{(k)}=E^{(k-1)}-F_{ij}^{(k-1)}
	\begin{bmatrix}
	0 & \Delta({i,j})\\
	\Delta({i,j})^T  & 0
	\end{bmatrix}.
	\end{align}
	Now, we consider the smallest eigenvalue of  matrix $E^{(k)}$ as a measure of distance to the boundary of constraint \eqref{Schured}. This is equal to the inverse of the largest eigenvalue of $(E^{(k)} )^{-1}$. Hence, we will keep track of matrix $(E^{(k)} )^{-1}$. It is straight-forward to show that \eqref{eq:firstupdate} is a rank-two update to matrix $E^{(k-1)}$ according to 
	\begin{align} \label{eq:E_eig}
	E^{(k)}=E^{(k-1)}+V(i,j)H^{(k-1)}(i,j)V^T(i,j),
	\end{align}
	where we have used the matrix of eigenvectors 
	\begin{align}
	V(i,j):=[v_1(i,j)\,|\, v_2(i,j)],
	\end{align}
	where $v_1(i,j) \in \mathbb{R}^{m+n}$ is a vector with $1/\sqrt{2}$ at locations $i$ and  $j+m$ and zero elsewhere. Also, $v_2(i,j) \in \mathbb{R}^{m+n}$ a vector with $1/\sqrt{2}$ at location $i$, and $-1/\sqrt{2}$ at location  $j+m$, and zero elsewhere. Moreover, $H^{(k-1)}(i,j)$ is given by 
	\begin{align}
	H^{(k-1)}(i,j):=\mathrm{diag}\left (-F_{ij}^{(k-1)} ,F_{ij}^{(k-1)} \right )
	\end{align}
	We can see that $v_1$ is perpendicular to $v_2$. Let us derive the update law for $(E^{(k)} )^{-1}$ as well.  The rank-two update derived for  $E^{(k)}$ can be used in Woodbury matrix identity \cite{woodbury1950inverting} to get
	\begin{align}\label{eq:invupdate}
	& (E^{(k)} )^{-1}=  (E^{(k-1)} )^{-1} -(E^{(k-1)} )^{-1} V(i,j) \times \notag \\
	& \left(\left(H^{(k-1)}(i,j)\right )^{-1} +V(i,j)^T(E^{(k-1)} )^{-1} V(i,j) \right )^{-1} \times \notag \\ 
	& V(i,j)^T(E^{(k-1)} )^{-1}.
	\end{align} 
	Then, we compute the maximum eigenvalue corresponding to each candidate for update using the Power method \cite{saad2011numerical}. Finally, at iteration $k$ we choose the $i_k$ and $j_k$ using  
	\begin{align}\label{eq:greedylaw}
	(i_k,j_k) \,\leftarrow ~ \underset{i,j}{\textrm{argmin}} \left \{ \lambda_{\max} \left ((E^{(k)} )^{-1}\right ):~ F^{(k-1)}_{ij} \neq 0 \right  \}
	\end{align} 
	and set the element at row $i_k$ and column $j_k$, (i.e, element $F_{i_kj_k}$) to zero. 
	Using this method as the sparsification tool in Algorithm $1$, the overall procedure would look like this:
	
	\begin{samepage}
		\noindent{\rule{8.65cm}{0.8pt}} \\ 
		{\bf {Algorithm 3:} \\ Greedy Sparsification inside Matrix Ellipsoid }
		\begin{enumerate}
			\item Start with $F^{(0)}=F_o$
			\item For $k=1,2,\dots $, choose $i$ and $j$ from 
			$$
			(i_k,j_k) \,\leftarrow~ \underset{i,j}{\textrm{argmin}} \left \{ \lambda_{\max} \left ((E^{(k)} )^{-1}\right ):~ F^{(k-1)}_{ij} \neq 0 \right  \},
			$$ 
			and find $F^{(k)}$ with setting $F_{i_kj_k}^{(k-1)}$ to zero in $F^{(k-1)}$. Stop when no such update exists. 
		\end{enumerate}
		\noindent{\rule{8.65cm}{0.8pt}}
	\end{samepage}
	
	\section{Numerical Examples} \label{NuExp}
	
	We examine the effectiveness of our proposed two-level algorithm. We define the following measures that are useful for assessment of the output of the sparsification. For a sparse controller $F$ that is built by searching around the pre-designed $F_o$, we define the density level of $F$ relative to $F_o$ (in percent) as 
	\begin{equation}
	\sigma_d  := 100 \times \dfrac{\|F\|_{0}}{\|F_o\|_{0}},
	\end{equation}
	and relative $\mathcal{H}_\infty$ performance loss (in percent) as 
	\begin{equation}
	\sigma_p := 100 \times \frac{\|G_F(s)\|_{\mathcal{H}_{\infty}}-\|G_{F_o}(s)\|_{\mathcal{H}_{\infty}}}{\|G_{F_o}(s)\|_{\mathcal{H}_{\infty}}},
	\end{equation}
	where  $G_F$ is the closed-loop transfer matrix from the disturbance to the output upon application of controller $F$. 
	
	\subsection{Sample Feedback Control Designs}
	
	We consider two classes of control systems and examine the effectiveness of our proposed procedure  as follows:
	
	\subsubsection{Randomly Generated System}  
\begin{figure}[t]
		\centering
		\vspace{.2cm}
		{\includegraphics[width=8.2cm]{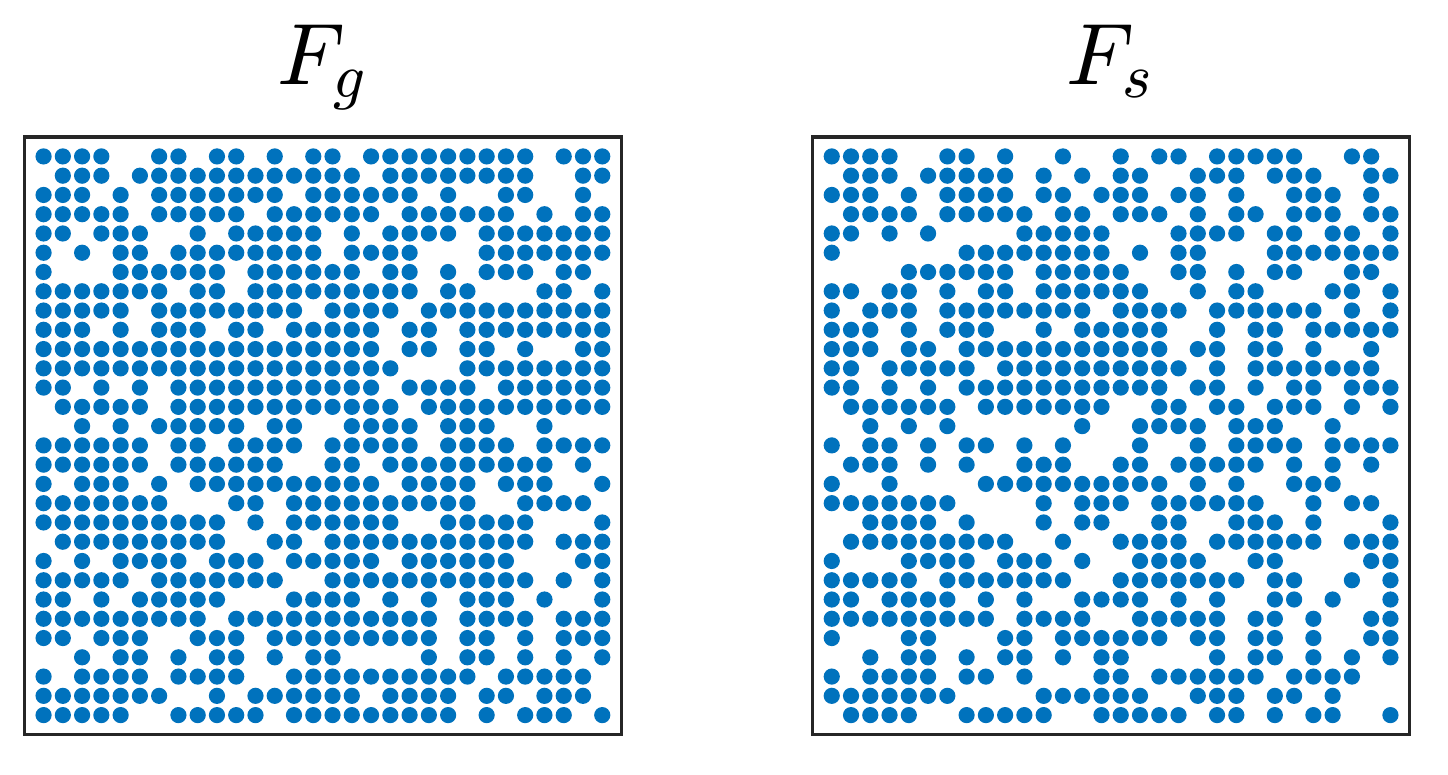}
		}
		\caption{\small The sparsity patterns of $F_s$ and $F_g$ for a uniform random system. Each blue dot corresponds to a nonzero element of the matrix.}
		\label{Figur1}
	\end{figure}

we randomly create the system matrix $A \in \R^{n\times n}$ and input matrix $B \in \R^{n\times m}$, where $n=m=30$ and each element of these matrices is independently  sampled from the standard normal distribution. We set the rest of the matrices to be $C=I$, $D_{gv}=I$, $D_{gu}=I$, and $B_v=B$.  First, we use the weighted $\ell_1$ regularization method with performance demand of  $\gamma=2$ and design parameter $\theta=0.5$ and find a value of feedback gain, which is denoted by $F_s \in \R^{m\times n}$. On the other hand, we use the greedy method with the same value of $\theta$ to find another feedback gain, namely, $F_g \in \R^{m\times n}$. The density level and performance measure of these controllers are shown in Table \ref{Tabll1}. It turns out that sparsity level $\sigma_d$ in $F_s$ and $F_g$ have decreased to about $65\%$ and $78\%$, while the relative performance degradation compared to the design with controller $F_o$ is only $7.4\%$ and $4.9\%$, respectively.  The sparsity patterns of these feedback gain designs have been illustrated in Fig. \ref{Figur1}.

	\subsubsection{Spatially-Decaying Interacting Subsystems:} 
\begin{figure}[t]
		\centering
		{\includegraphics[width=8.5cm]{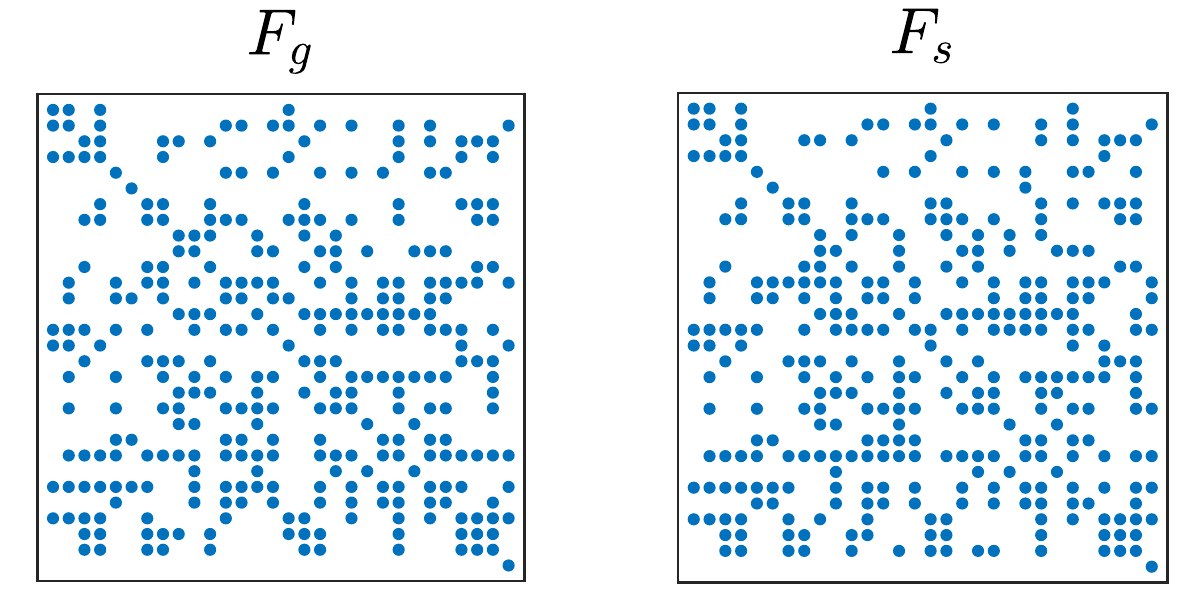}
		}
		\caption{\small The sparsity patterns of feedback gains $F_s$ and $F_g$ for the exponentially decaying random system. Each blue dot corresponds to a nonzero element of the matrix.}
		\label{Figur2}
	\end{figure}

\begin{figure}[t]
		\centering
		\vspace{.2cm}
		{\includegraphics[width=7cm]{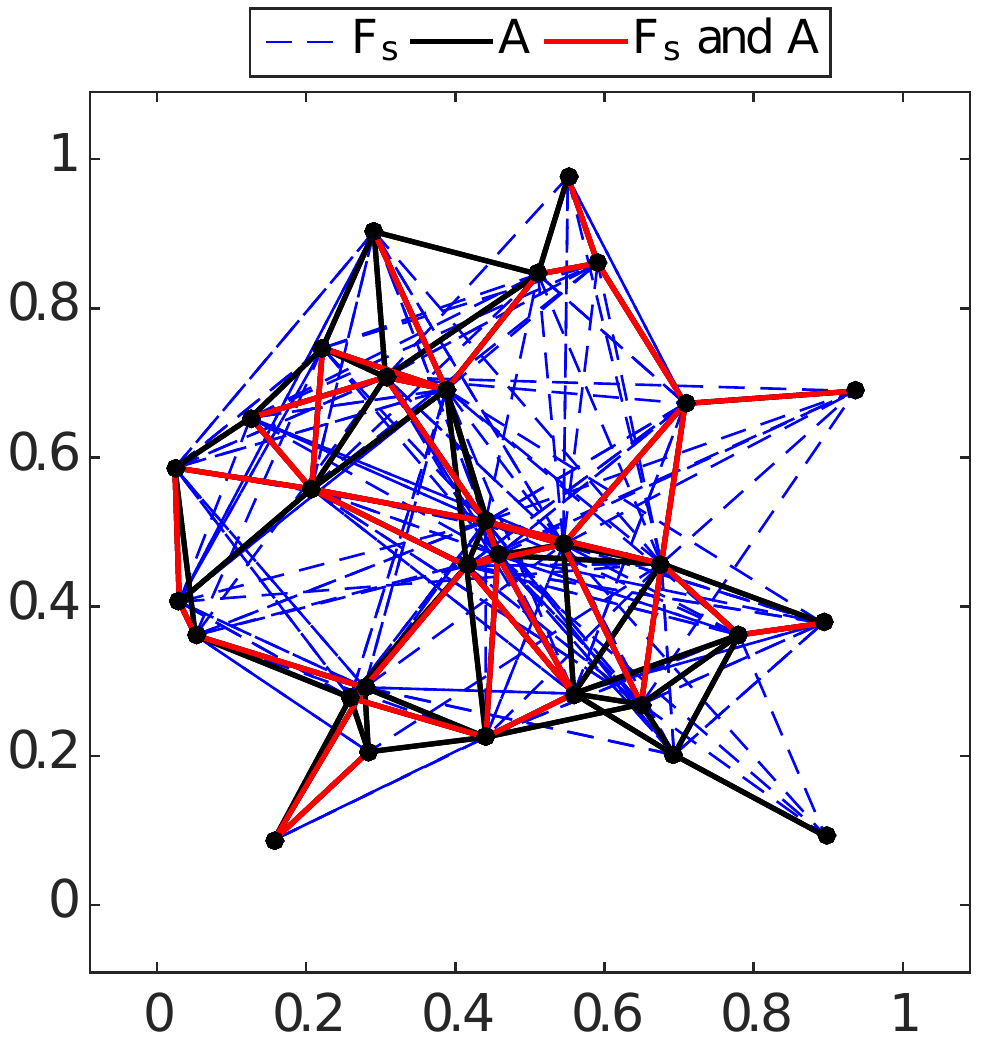}
		}
		\caption{\small The links in the sparse controller $F_s$ and the system with spatially decaying interconnections. The links that exist in both structures have been highlighted as well. }
		\label{fig:topology}
	\end{figure}

	We consider the matrix $A$ to be constructed as follows: We consider $n$ agents that are located in the square-shaped region $[0,1]\times [0,1]$, with a random position $p_l \in [0,1] \times [0,1]$ for the $l$'th agent that was uniformly picked. 
	Then, we suppose that the element in row $i$ and column $j$ of $A$, namely, $A_{ij}$ is zero if $\|p_i-p_j\|_2> r$, and otherwise its value is given by
	$$
	A_{ij}=c_{ij} \mathrm{e}^{-\alpha \|p_i-p_j\|_2^\beta},
	$$
	where ${c}_{ij} \sim \mathcal{N}(0,1)$, $\alpha$ determines the bandwidth of the system,   $\beta$ specifies the decay rate of the interaction between two spatially located agents, and $r$ is the radius of the connectivity disk around each agent. We consider  $n=30$ with system parameters $\alpha=\beta=1$ and ${r=0.25}$.
	We set $B=I$, (i.e., each agent has an actuator that controls its state), and $C=D_{gv}=D_{gu}=I$, $B_v= 4 B$.
	For the sparsification parameters, we  
	set $\gamma=5$, $\theta=0.5$, 
	and similar to the previous example, we obtain two sparse controllers $F_s$ and $F_g$ using the weighted $\ell_1$ regularization and greedy method, respectively. Their corresponding sparsity patterns are visualized in Fig. \ref{Figur2}. According to Table \ref{Tabll2}, the full density level of $F_o$ has decreased to almost $37.5\%$ and $36.3\%$, while the performance loss is around $1.2\%$ and $0.4\%$, 
	in the case of $F_s$ and $F_g$, respectively. In Fig. \ref{fig:topology}, we have illustrated the links corresponding to connectivity architecture of the system as well as the links that correspond to the information structure of sparse controller $F_s$. Higher levels of achieved sparsity in this example are supposedly due to the diagonal input matrix $B$. 
	
	\begin{table}[t]
		\vspace{.2cm}
		\small
		\centering
		\begin{tabular}{|c|c|c|c|} 
			\hline 
			$F$ & $F_o$ & $F_s$& $F_g$ \\ \hline
			$\|F\|_0$ & $900$ & $587$ & $701$ \\ \hline 
			$\|G_F(s)\|_{\mathcal{H}_\infty}$ & $1.0001$ & $1.0740$  & $1.0487$ \\ \hline
		\end{tabular}
		\caption{\small $\mathcal{H}_{\infty}$ norm and number of nonzero elements in $F_o$, $F_s$, and $F_g$ (for a randomly generated system).}
		\normalsize
		\label{Tabll1}
	\end{table}
	
	\begin{table}[t]
		\small
		\centering
		\begin{tabular}{|c|c|c|c|} 
			\hline 
			$F$ & $F_o$ & $F_s$& $F_g$ \\ \hline
			$\|F\|_0$ & $900$ & $337$ & $326$ \\ \hline 
			$\|G_F(s)\|_{\mathcal{H}_\infty}$ & $3.4574$ & $3.4993$  & $3.4723$ \\ \hline
		\end{tabular}
		\caption{\small $\mathcal{H}_{\infty}$ norm and number of nonzero elements in $F_o$, $F_s$, and $F_g$ (for a sub-exponentially decaying randomly generated system).}
		\normalsize
		\label{Tabll2}
	\end{table}
	
	\begin{figure}[t]
		\centering
		{\includegraphics[width=8cm]{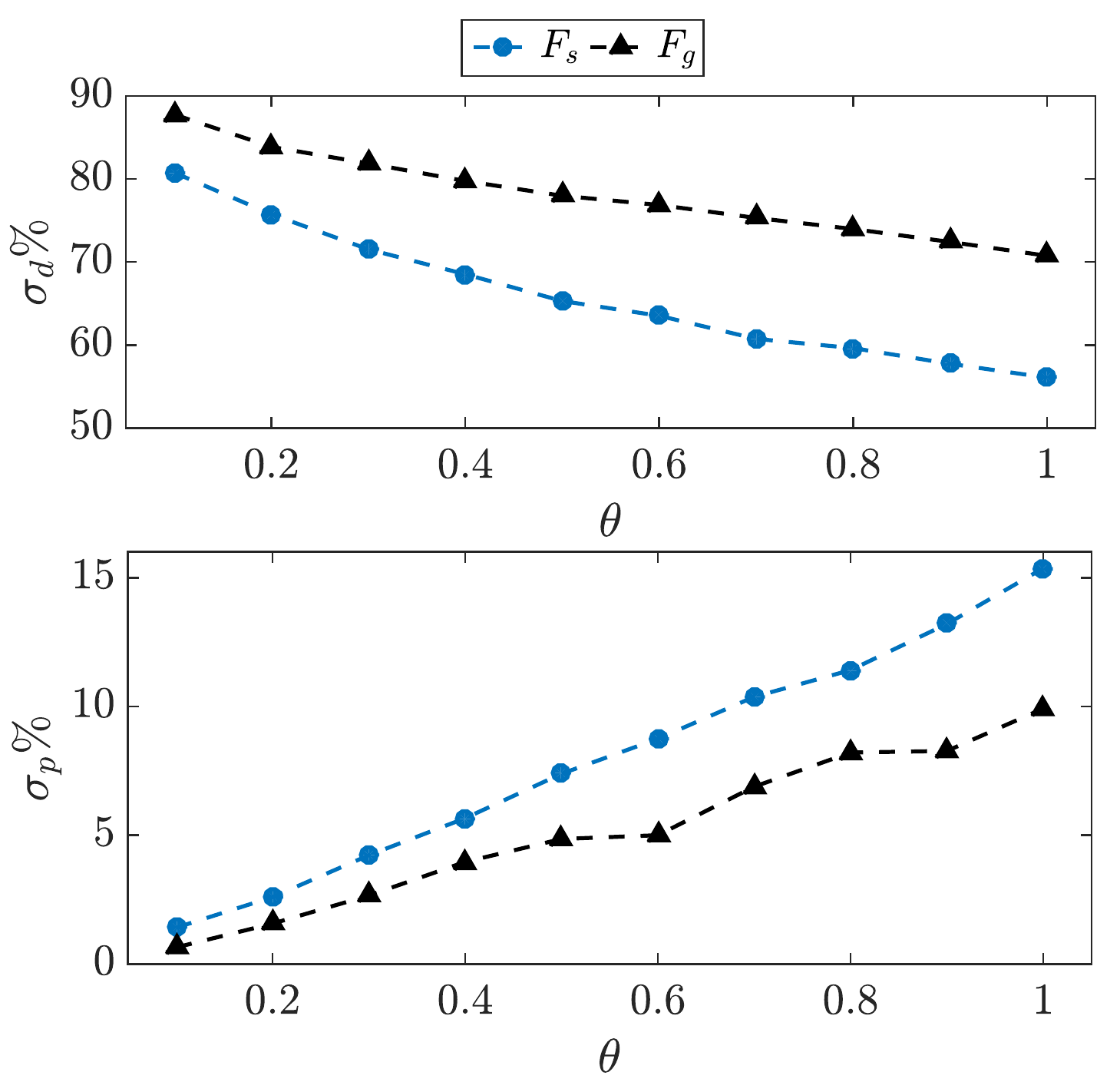}
		}
		\caption{\small Relative density levels and relative performance losses of sparse controllers $F_s$ and $F_g$ vs the parameter $\theta$. Recall that $F_s$ and $F_g$ are outputs of sparsification using Algorithms 2 and 3, respectively.}
		\label{fig:tradeoffs_rand}
	\end{figure}
	
	\begin{figure}[t]
		\centering
		\vspace{.2cm}
		{\includegraphics[width=8cm]{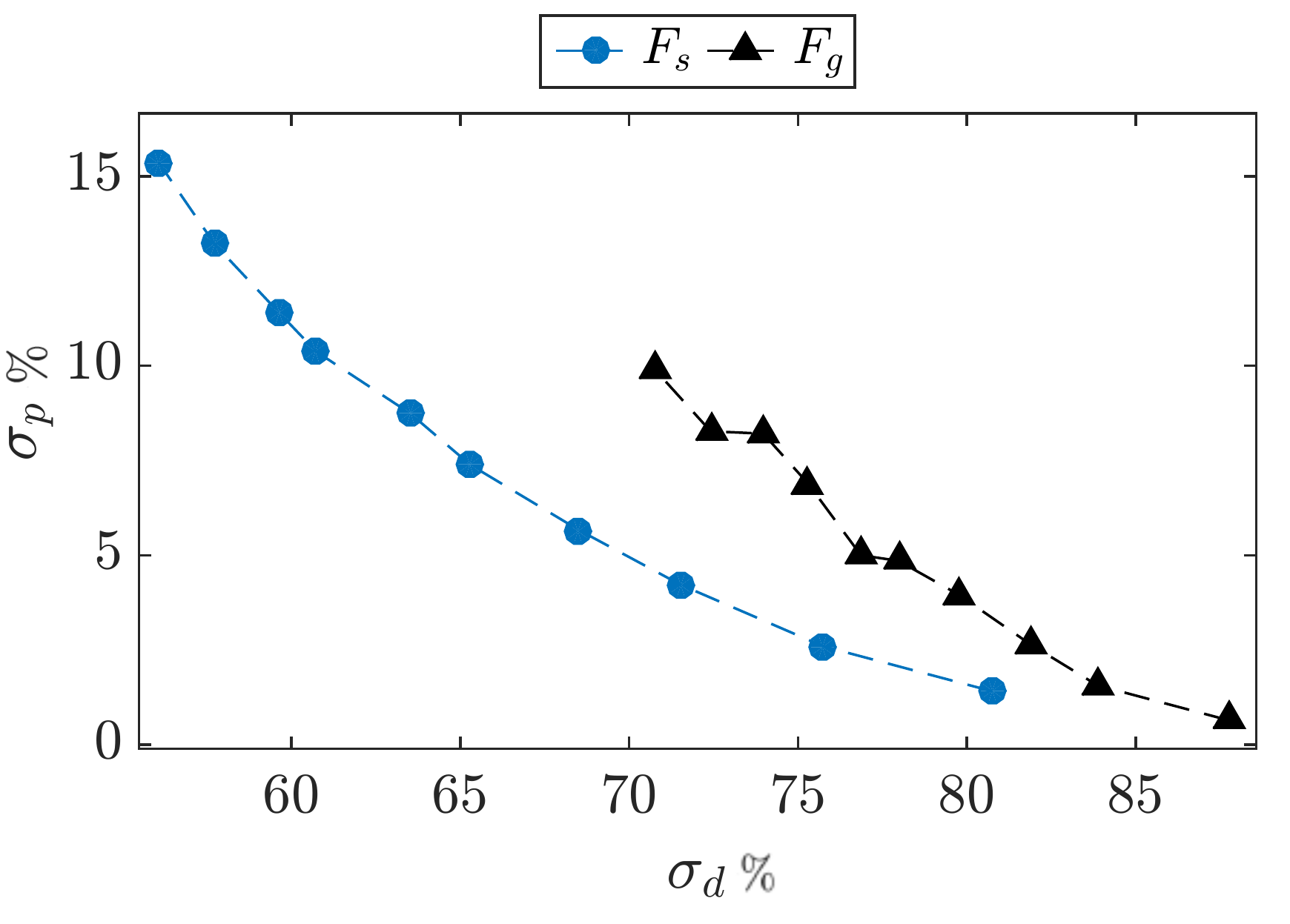}
		}
		\caption{\small The tradeoff curves between the density levels and the relative performance degradation for sparse set of controllers $F_s$ and $F_g$.}
		\label{fig:tradeoffs_curve}
	\end{figure}
	
	\subsection{Interplay between Resilience, Performance, and Sparsity}
	\subsubsection*{Randomly Generated System}
	Reconsidering the control system given in the previous subsection, we vary parameter $\theta$ from $0$ to $1$ and look at the resulting density levels and performance losses of the sparse controllers.  In Fig. \ref{fig:tradeoffs_rand}, we observe that as the parameter $\theta$ increases, the density level of the outcome is enhanced, while the performance measure is deteriorating.   We observe that these data represent a tradeoff between the relative density level of the feedback gain and the relative performance loss upon sparsification, which is further illustrated in Fig. \ref{fig:tradeoffs_curve}. In these examples, the greedy method turns out to be more conservative, while trading more performance for lower levels of sparsity.
	
	If we do the same experiment on the control system with spatially decaying parameter, we arrive at similar trends and tradeoffs, which we omit for brevity.  
	
	\subsection{Study of the Resilience}
	
	We create a random sample of the system with sub-exponentially decaying interactions with $n=20$, $\alpha=1$, $\beta=0.5$, and {$r=0.6$}. Then, we study the resilience of the designed sparse feedback control laws using re-weighted $\ell_1$ regularization. We vary the parameter $\theta$ to find a sparse feedback gain. Then, given the sparse controller $F_s$, we consider random perturbations of each nonzero element of the controller that have the form
	$$
	[F_s]_{ij} \leftarrow [F_s]_{ij}+0.5\,\eta_{ij},
	$$
	where $\eta_{ij} \sim \mathcal{N}(0,1)$, independent of other elements. For each design, (i.e., each value of the sparse controller corresponding to the value of $\theta$), we find $5,000$ random perturbations. Then, for each randomly perturbed feedback gain, we evaluate the $\mathcal{H}_\infty$ performance measure of the closed-loop system.
	
\begin{figure}[t]
		\centering
		{\includegraphics[width=8cm]{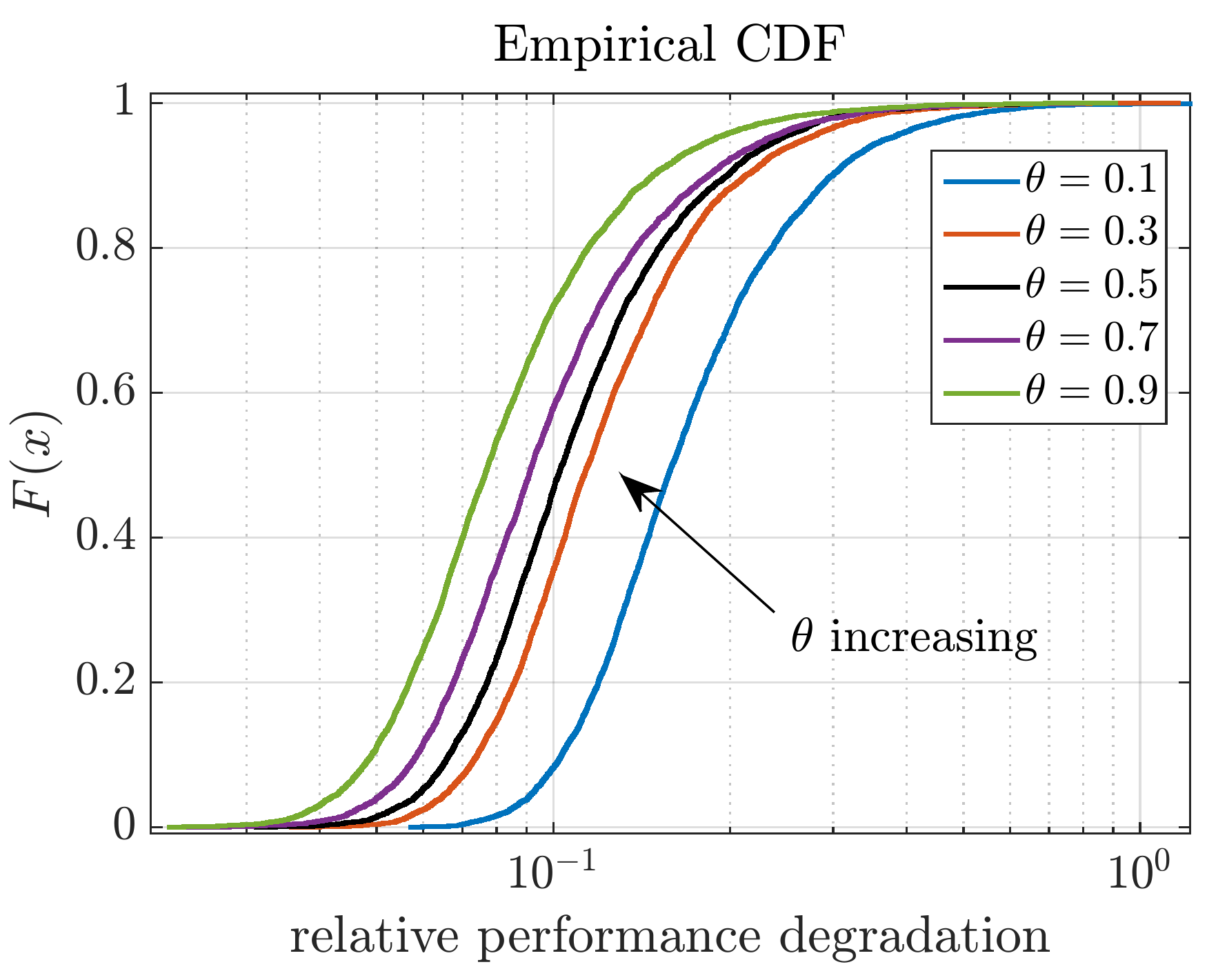}
		}
		\caption{\small The empirical CDF of the relative degradation in the $\mathcal{H}_\infty$ performance measure $\sigma_p$ for 5,000 random perturbations of the controller with different tuning parameter $\theta$.}
		\label{fig:cdfplot}
	\end{figure}

	The empirical cumulative density function (CDF) of the degradation in the $\mathcal{H}_\infty$ performance measure (relative the sparse design) under these random perturbations is illustrated in Fig. \ref{fig:cdfplot}. We observe that for the larger values of $\theta$, the sparse control design is more fragile. 
	
	\section{Final Remarks} \label{DandC}
	
	\noindent 1) The general methodology followed in this paper can be summarized as follows: first, obtain a region in the controller space inside which a specific control objective holds. Then, explore the geometry of the region in the controller space to find a sparse controller that inherits the performance guarantee. While in this paper we have addressed $\mathcal{H}_\infty$ control design using the quadratic inequalities, as mentioned in \cite{peaucelle2004lmi}, one can do similar developments for the $\mathcal{H}_2$ norm performance measure as well.
	
	\noindent 2) Because some of the matrix inequalities involved in the first method are strict, solving the linear matrix inequalities in the first stage is subject to a great deal of flexibility. For instance, from extensive numerical simulations, we have learned that limiting the condition number of matrix $P$ in \eqref{LMS} is practical for finding a  ''rich'' enough region of controllers that will be used in the sparsification. 
	
	\noindent 3) Although we have limited the value of $\theta$ to $[0,1]$, some numerical examples suggest that even for values of $\theta$ greater than $1$ the second step for finding sparse controllers may result in satisfactory results. This is due to the conservative nature of the quadratic regions of controllers \cite{peaucelle2004lmi}. 
	
	\noindent 4) The exact running time analysis for the sparsification methods may not be derivable. However, we expect that the greedy method becomes less computationally expensive compared to $\ell_1$ regularization as the size of the system increases.
	
	\noindent 5) Our extensive numerical studies suggest that as the designed controller become sparser, the fragility of the design would increase as well.
	
	\noindent 6) The distance measure used for greedy sparsification, (i.e., smallest eigenvalue) can be also changed to a number of other measures. For instance, we can replace the protocol \eqref{eq:greedylaw}  to be   
		$$
		(i_k,j_k) \,\leftarrow ~ \underset{i,j}{\textrm{argmin}} \left \{ \sum_{i=1}^{m+n} \lambda_{i}\left ((E^{(k)} )^{-1}\right ):~ F^{(k-1)}_{ij} \neq 0 \right  \},
		$$
		where $\lambda_i$ denotes the $i$'th eigenvalue of the matrix.  Alternatively, one can set
		$$
		(i_k,j_k) \,\leftarrow ~ \underset{i,j}{\textrm{argmin}} \left \{  \log \left  (\det\left ((E^{(k)} )^{-1} \right )\right ):~ F^{(k-1)}_{ij} \neq 0 \right  \}.
		$$
		These measures capture alternative aspects of distance of the matrix from the boundary of positive-definite matrices and we can derive simple update laws  for these measures using update formula \eqref{eq:invupdate} (e.g. see \cite{mousavi2018space}). 
	
	\bibliographystyle{IEEEtran}
	
	\bibliography{SPFRAG}

\begin{thebibliography}{10}
\providecommand{\url}[1]{#1}
\csname url@samestyle\endcsname
\providecommand{\newblock}{\relax}
\providecommand{\bibinfo}[2]{#2}
\providecommand{\BIBentrySTDinterwordspacing}{\spaceskip=0pt\relax}
\providecommand{\BIBentryALTinterwordstretchfactor}{4}
\providecommand{\BIBentryALTinterwordspacing}{\spaceskip=\fontdimen2\font plus
\BIBentryALTinterwordstretchfactor\fontdimen3\font minus
  \fontdimen4\font\relax}
\providecommand{\BIBforeignlanguage}[2]{{%
\expandafter\ifx\csname l@#1\endcsname\relax
\typeout{** WARNING: IEEEtran.bst: No hyphenation pattern has been}%
\typeout{** loaded for the language `#1'. Using the pattern for}%
\typeout{** the default language instead.}%
\else
\language=\csname l@#1\endcsname
\fi
#2}}
\providecommand{\BIBdecl}{\relax}
\BIBdecl

\bibitem{takahashi2000robust}
R.~H. Takahashi, D.~A. Dutra, R.~M. Palhares, and P.~L. Peres, ``On robust
  non-fragile static state-feedback controller synthesis,'' in
  \emph{Proceedings of the 39th IEEE Conference on Decision and Control},
  vol.~5, 2000, pp. 4909--4914.

\bibitem{famularo2000robust}
D.~Famularo, P.~Dorato, C.~T. Abdallah, W.~M. Haddad, and A.~Jadbabaie,
  ``Robust non-fragile lq controllers: the static state feedback case,''
  \emph{International Journal of control}, vol.~73, no.~2, pp. 159--165, 2000.

\bibitem{park2004robust}
J.~H. Park, ``Robust non-fragile control for uncertain discrete-delay
  large-scale systems with a class of controller gain variations,''
  \emph{Applied Mathematics and Computation}, vol. 149, no.~1, pp. 147--164,
  2004.

\bibitem{peaucelle2005ellipsoidal}
D.~Peaucelle and D.~Arzelier, ``Ellipsoidal sets for resilient and robust
  static output-feedback,'' \emph{IEEE Transactions on Automatic Control},
  vol.~50, no.~6, pp. 899--904, 2005.

\bibitem{rotkowitz2006characterization}
M.~Rotkowitz and S.~Lall, ``A characterization of convex problems in
  decentralized control,'' \emph{IEEE Transactions on Automatic Control},
  vol.~51, no.~2, pp. 274--286, 2006.

\bibitem{motee2008optimal}
N.~Motee and A.~Jadbabaie, ``Optimal control of spatially distributed
  systems,'' \emph{IEEE Transactions on Automatic Control}, vol.~53, no.~7, pp.
  1616--1629, 2008.

\bibitem{lin2013design}
F.~Lin, M.~Fardad, and M.~R. Jovanovi{\'c}, ``Design of optimal sparse feedback
  gains via the alternating direction method of multipliers,'' \emph{IEEE
  Transactions on Automatic Control}, vol.~58, no.~9, pp. 2426--2431, 2013.

\bibitem{dorfler2014sparsity}
F.~D{\"o}rfler, M.~R. Jovanovi{\'c}, M.~Chertkov, and F.~Bullo,
  ``Sparsity-promoting optimal wide-area control of power networks,''
  \emph{IEEE Transactions on Power Systems}, vol.~29, no.~5, pp. 2281--2291,
  2014.

\bibitem{dhingra2016method}
N.~K. Dhingra and M.~R. Jovanovi{\'c}, ``A method of multipliers algorithm for
  sparsity-promoting optimal control,'' in \emph{2016 American Control
  Conference (ACC)}.\hskip 1em plus 0.5em minus 0.4em\relax IEEE, 2016, pp.
  1942--1947.

\bibitem{arastoo2016closed}
R.~Arastoo, M.~Bahavarnia, M.~V. Kothare, and N.~Motee, ``Closed-loop feedback
  sparsification under parametric uncertainties,'' in \emph{IEEE 55th
  Conference on Decision and Control (CDC)}, 2016, pp. 123--128.

\bibitem{bahavarnia2017state}
M.~Bahavarnia, C.~Somarakis, and N.~Motee, ``State feedback controller
  sparsification via a notion of non-fragility,'' in \emph{IEEE 56th Annual
  Conference on Decision and Control (CDC)}, 2017, pp. 4205--4210.

\bibitem{bahavarnia2019state}
M.~Bahavarnia, ``State-feedback controller sparsification via quasi-norms,'' in
  \emph{American Control Conference (ACC)}.\hskip 1em plus 0.5em minus
  0.4em\relax IEEE, 2019, pp. 748--753.

\bibitem{sojoudi2011structurally}
S.~Sojoudi, J.~Lavaei, and A.~G. Aghdam, \emph{Structurally Constrained
  Controllers: Analysis and Synthesis}.\hskip 1em plus 0.5em minus 0.4em\relax
  Springer Science \& Business Media, 2011.

\bibitem{lavaei2013optimal}
J.~Lavaei, ``Optimal decentralized control problem as a rank-constrained
  optimization,'' in \emph{2013 51st Annual Allerton Conference on
  Communication, Control, and Computing (Allerton)}.\hskip 1em plus 0.5em minus
  0.4em\relax IEEE, 2013, pp. 39--45.

\bibitem{sadabadi2014fixed}
M.~S. Sadabadi and A.~Karimi, ``Fixed-structure sparse control of
  interconnected systems with polytopic uncertainty,'' \emph{IFAC Proceedings
  Volumes}, vol.~47, no.~3, pp. 2588--2593, 2014.

\bibitem{wang2014sparse}
Y.~Wang, J.~Lopez, and M.~Sznaier, ``Sparse static output feedback controller
  design via convex optimization,'' in \emph{53rd IEEE Conference on Decision
  and Control}.\hskip 1em plus 0.5em minus 0.4em\relax IEEE, 2014, pp.
  376--381.

\bibitem{fattahi2017convexity}
S.~Fattahi and J.~Lavaei, ``On the convexity of optimal decentralized control
  problem and sparsity path,'' in \emph{2017 American Control Conference
  (ACC)}.\hskip 1em plus 0.5em minus 0.4em\relax IEEE, 2017, pp. 3359--3366.

\bibitem{wu2017sparsity}
X.~Wu and M.~R. Jovanovi{\'c}, ``Sparsity-promoting optimal control of systems
  with symmetries, consensus and synchronization networks,'' \emph{Systems \&
  Control Letters}, vol. 103, pp. 1--8, 2017.

\bibitem{jovanovic2014sparsity}
M.~R. Jovanovi{\'c}, P.~J. Schmid, and J.~W. Nichols, ``Sparsity-promoting
  dynamic mode decomposition,'' \emph{Physics of Fluids}, vol.~26, no.~2, p.
  024103, 2014.

\bibitem{yu2018structured}
C.~Yu and M.~Verhaegen, ``Structured modeling and control of adaptive optics
  systems,'' \emph{IEEE Transactions on Control Systems Technology}, vol.~26,
  no.~2, pp. 664--674, 2018.

\bibitem{peaucelle2004lmi}
D.~Peaucelle, D.~Arzelier, and C.~Farges, ``Lmi results for resilient
  state-feedback with $\mathcal{H}_\infty$ performance,'' in \emph{43rd IEEE
  Conference on Decision and Control (CDC)}, vol.~1, 2004, pp. 400--405.

\bibitem{zhou1996robust}
K.~Zhou, J.~C. Doyle, K.~Glover \emph{et~al.}, \emph{Robust and optimal
  control}, vol.~40.

\bibitem{mital2015cloud}
M.~Mital, A.~K. Pani, S.~Damodaran, and R.~Ramesh, ``Cloud based management and
  control system for smart communities: A practical case study,''
  \emph{Computers in Industry}, vol.~74, pp. 162--172, 2015.

\bibitem{blondel1997np}
V.~Blondel and J.~N. Tsitsiklis, ``Np-hardness of some linear control design
  problems,'' \emph{SIAM Journal on Control and Optimization}, vol.~35, no.~6,
  pp. 2118--2127, 1997.

\bibitem{grant2008cvx}
M.~Grant, S.~Boyd, and Y.~Ye, ``Cvx: Matlab software for disciplined convex
  programming,'' 2008.

\bibitem{zhang2006schur}
F.~Zhang, \emph{The Schur complement and its applications}.\hskip 1em plus
  0.5em minus 0.4em\relax Springer Science \& Business Media, 2006, vol.~4.

\bibitem{Candes:2008}
E.~Candes, M.~Wakin, and S.~Boyd, ``Enhancing sparsity by reweighted $\ell_1$
  minimization,'' \emph{Journal of Fourier Analysis and Applications}, vol.~14,
  pp. 877--905, July 2008.

\bibitem{woodbury1950inverting}
M.~A. Woodbury, ``Inverting modified matrices,'' \emph{Memorandum report},
  vol.~42, no. 106, p. 336, 1950.

\bibitem{saad2011numerical}
Y.~Saad, \emph{Numerical methods for large eigenvalue problems: revised
  edition}.\hskip 1em plus 0.5em minus 0.4em\relax Siam, 2011, vol.~66.

\bibitem{mousavi2018space}
H.~K. Mousavi, Q.~Sun, and N.~Motee, ``Space-time sampling for network
  observability,'' \emph{arXiv preprint arXiv:1811.01303}, 2018.

\end{thebibliography}
	
\end{document}